\documentclass[sigconf,natbib=true]{acmart}
\AtBeginDocument{%
  }

\setcopyright{acmlicensed}
\copyrightyear{2026}
\acmYear{2026}
\setcopyright{cc}
\setcctype{by}
\acmConference[SIGIR '26]{Proceedings of the 49th International ACM SIGIR Conference on Research and Development in Information Retrieval}{July 20--24, 2026}{Melbourne, VIC, Australia}
\acmBooktitle{Proceedings of the 49th International ACM SIGIR Conference on Research and Development in Information Retrieval (SIGIR '26), July 20--24, 2026, Melbourne, VIC, Australia}
\acmDOI{10.1145/3805712.3809667}
\acmISBN{979-8-4007-2599-9/2026/07}

\usepackage[table]{xcolor} 
\usepackage{makecell}
\usepackage{multirow}
\usepackage{subcaption}
\usepackage{todonotes}

\begin{document}

\title{How Generative AI Disrupts Search: An Empirical Study of Google Search, Gemini, and AI Overviews}

\author{Riley Grossman}
\orcid{0009-0009-1114-6375}
\affiliation{%
  \institution{New Jersey Institute of Technology}
  \city{Newark}
  \state{NJ}
  \country{USA}
}
\email{rag24@njit.edu}

\author{Songjiang Liu}
\orcid{0009-0006-6669-5951}
\affiliation{%
  \institution{New Jersey Institute of Technology}
  \city{Newark}
  \state{NJ}
  \country{USA}
}
\email{sl947@njit.edu}

\author{Michael K. Chen}
\orcid{0000-0002-2727-5798}
\affiliation{%
  \institution{Nanyang Technological University}
  \city{Singapore}
  \country{Singapore}
}
\email{michaelchenkj@gmail.com}

\author{Mike Smith}
\orcid{0000-0002-7649-353X}
\affiliation{%
 \institution{Indiana University Bloomington}
 \city{Bloomington}
 \state{IN}
 \country{USA}}
\email{ms255@iu.edu}

\author{Cristian Borcea}
\orcid{0000-0003-0020-0910}
\affiliation{%
  \institution{New Jersey Institute of Technology}
  \city{Newark}
  \state{NJ}
  \country{USA}}
\email{borcea@njit.edu}

\author{Yi Chen}
\orcid{0000-0003-3669-1643}
\affiliation{%
  \institution{New Jersey Institute of Technology}
  \city{Newark}
  \state{NJ}
  \country{USA}}
\email{yi.chen@njit.edu}

\renewcommand{\shortauthors}{Grossman et al.}


\newcommand{\yc}[1]{\textcolor{red}{}}

\newcommand{\newtexts}[1]{\textcolor{blue}{#1}}

\begin{abstract}

Generative AI is being increasingly integrated into web search for the convenience it provides users. In this work, we aim to understand how generative AI disrupts web search by retrieving and presenting the information and sources differently from traditional search engines. 
We introduce a public benchmark dataset of 11,500 user queries to support our study and future research of generative search. We compare the search results returned by Google's search engine, the accompanying AI Overview (AIO), and Gemini Flash 2.5 for each query. We have made several key findings. First, we find that for 51.5\% of representative, real-user queries, AIOs are generated, and are displayed above the organic search results. Controversial questions frequently result in an AIO. Second, we show that the retrieved sources are substantially different for each search engine (<0.2 average Jaccard similarity). Traditional Google search is significantly more likely to retrieve information from popular or institutional websites in government or education, while generative search engines are significantly more likely to retrieve Google-owned content. Third, we observe that websites that block Google's AI crawler are significantly less likely to be retrieved by AIOs, despite having access to the content. Finally, AIOs are less consistent when processing two runs of the same query, and are less robust to minor query edits. Our findings have important implications for understanding how generative search impacts website visibility, the effectiveness of generative engine optimization techniques, and the information users receive. We call for revenue frameworks to foster a sustainable and mutually beneficial ecosystem for publishers and generative search providers.

\end{abstract}




\newcommand{\michael}[1]{\textcolor{blue}{#1}}
\newcommand{\ryan}[1]{\textcolor{orange}{#1}}

\maketitle

\section{Introduction}







\begin{figure}
    \centering
    \includegraphics[width=\linewidth]{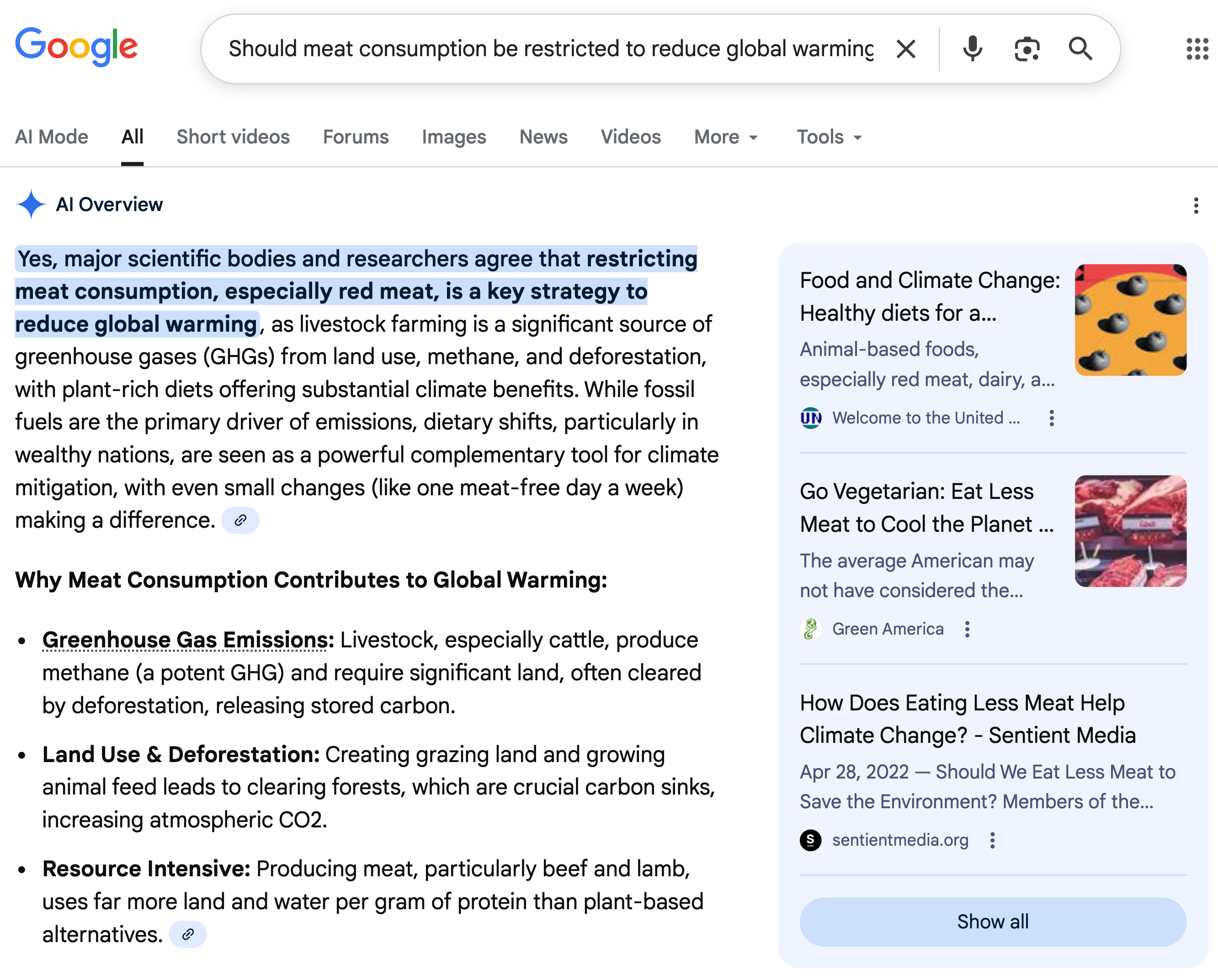}
    \caption{Google's AI Overview Interface}
    \label{fig:aio_interface}
\end{figure}

With the rapid advancement of large language models (LLMs), users have begun to use AI chatbots (e.g., ChatGPT) as a replacement for traditional search engines~\cite{demand_sage_gpt,first_page_sage,vis_capitalist}. In response, traditional search engines have integrated LLM capabilities into their search functionalities, thereby affecting traditional search engine users in a way that AI chatbots do not. For example, while users must choose to use AI chatbots, Google places its AI Overview (AIO) above the traditional search engine results pages (SERP) by default (see Figure~\ref{fig:aio_interface}).\footnote{Users can temporarily remove AIOs by adding ``-AI'' to a query or scrolling to the ``Web'' tab in the search results.} We use the term generative search engines to refer to any AI-powered systems that retrieve relevant sources in response to a user query and generate a summary of the content in those sources.

Generative search engines offer convenience to users~\cite{liang_how_2025,stadler_cognitive_2024,dai_next-search_2025} and are rapidly gaining popularity. However, it remains unclear how the sources retrieved by generative search engines compare with those returned by traditional search, what implications such differences have for users, and how these differences affect websites. 

Users rely on the retrieved sources to obtain information, particularly those ranked highly by search engines~\cite{baye2016search,seo_ieee_paper,zilincan2015search}. Thus, it is critical that search engines consistently retrieve quality sources (e.g., for supporting an informed electorate in democratic society~\cite{dai_bias_2024,bozdag2015breaking,steiner2022seek}). Furthermore, for usability, it is important that search engines retrieve consistent results in the presence of minor changes to the query syntax that preserve intent~\cite{bailey2017retrieval,hannak2013measuring,jiang2015improving,rank_stability}.

From a website's perspective, there is growing concern that generative search reduces website traffic, as users increasingly obtain information directly from generated summaries rather than from sources listed in traditional SERPs~\cite{seosandwitch}. Historically, websites have relied on Search Engine Optimization (SEO) services to improve their rankings in traditional search results in order to increase traffic, thereby increase advertising revenue or sales opportunities. More recently, websites have begun exploring strategies to offset declining traffic by increasing their presence as cited sources in generative search results~\cite{seer,submitshop}. Although some companies offer Generative Engine Optimization (GEO) services to increase website visibility in generative search results, with some evidence of its potential~\cite{aggarwal_geo_2024,pfrommer-etal-2024-ranking,chen2025products}, the effectiveness of GEO is contested~\cite{puerto2025cseo}. More fundamentally, there is a lack of understanding of how the characteristics of sources cited by generative search differ from those retrieved by traditional search engines.

In this paper, we conduct a large-scale comparative analysis of the sources retrieved by traditional and generative search engines. We construct a benchmark query set of 11,500 queries, along with two additional time-sensitive query sets, spanning diverse topics, user intents, and syntactic variations. Using SerpAPI\footnote{\label{fn:serpapi}\url{https://github.com/serpapi}} and the Gemini API, we collect ranked lists of retrieved sources from Google Search (i.e., traditional SERP), AIOs, and Gemini 2.5 Flash. We then use Jaccard similarity and rank-biased overlap (RBO) to quantify differences between the sources retrieved by each engine, and also to evaluate each system’s internal consistency across repeated runs and in response to variations in query syntax, user location, and device. We also collect domain-level characteristics (e.g., popularity and content category) of the retrieved sources to assess the source preferences of each search engine.
%

%
%

We summarize our findings as follows:%
\begin{itemize}
    \item  AIOs have a substantial presence (e.g., 51.5\% of representative real-user queries), particularly in response to long, informational queries formatted as a question.    
    \item Traditional SERP, Gemini, and AIO exhibit low average similarity in terms of the retrieved sources for each query (i.e., Jaccard similarities between 0.11 and 0.18). 
    \item Generative search engines are significantly less likely to retrieve sources from popular websites, educational or governmental institutions, and websites that block Google's AI bot (even though AIOs can still access the content). Generative search engines retrieve significantly more content from Google-owned websites. 
    \item Generative search is less consistent across runs of the same query, and less robust to changes in device type or query syntax, compared to traditional search.
    \item Although AIOs are rarely generated for trending queries (8.1\%), they are prevalent for sensitive queries (e.g., 93.8\% of political queries). Despite the importance of these queries to maintaining an informed electorate, generative search engines rely on less credible sources. They also often take a stance in the generated text summary (33.4\% of AIO, and 5.6\% of Gemini, responses, respectively).
   \end{itemize}

   Our findings have important societal implications. Users should exercise caution when using generative AI due to the risks of inaccuracies and hallucinations, and regulation may be necessary for controversial and high-stakes queries. Our results also inform publishers’ use of GEO and decisions about blocking AI crawling. We highlight several recommendations for improvements in the generative search ecosystem. The current dilemma is that reputable publishers often restrict AI crawler access, leading to further traffic  reductions, while degrading the quality of generative search results. We call for revenue frameworks that align incentives between publishers and generative search providers to foster a sustainable and mutually beneficial ecosystem.
%

To enable future research on the evolution and sustainability of generative search, processed datasets and code are available at: \url{https://github.com/rag24/AIO}.
\section{Related Work}

Generative search engines and AI-based information retrieval (IR) systems have become increasingly popular due to the convenience and user satisfaction they provide~\cite{liang_how_2025,stadler_cognitive_2024,dai_next-search_2025}. Prior studies have examined the quality of generated summaries and identified issues such as bias and hallucination~\cite{dai_bias_2024,dai_trustworthy_2025,zhang_large_2024,stadler_cognitive_2024,ben-tov_gasliteing_2025}.
User studies have shown that generative search engines may decrease the quality of information users obtain~\cite{stadler_cognitive_2024} and create echo-chambers~\cite{sharma,narayanan_venkit_search_2025}. Studies have found that 25\% of generative search citations do not support the corresponding sentence~\cite{liu-etal-2023-evaluating}, and information is often cherry-picked or attributed to the wrong source~\cite{narayanan_venkit_search_2025}. 
Different from these works, which analyze the quality of the generated summary, our study compares traditional and generative search engines, specifically through analyzing the differences in each engine's retrieved sources.

There are several studies that evaluate offline AI-based IR systems, where the AI system can only access a limited number of documents for each query~\cite{dai_mitigating_2025,dai_neural_2024,xu_images,dai-etal-2025-media}. 
Several studies show that AI-based IR systems favor information from questionable sources, such as AI generated content ~\cite{dai_mitigating_2025,dai_neural_2024,xu_images} and politically biased content~\cite{dai-etal-2025-media}. We also study the quality of sources retrieved by AI systems, but we evaluate online generative search engines that access the entire Web in response to real user queries, rather than in an offline limited document setting.

As an important area that is gaining increasing attention, three relevant preprint studies have recently emerged.
One study compares generative search engines' differing reliance on internal knowledge contained within the underlying LLM versus external knowledge from the Web~\cite{kirsten2025}. The other two studies also evaluate differences between the sources cited by generative search engines and those retrieved by traditional search, but they exclude AIOs and are conducted on a very small set of specialized queries: one study focuses on  product and service recommendation queries~\cite{chen2025products}, and the other evaluates only 24 sociopolitical queries~\cite{minici}. In contrast, our paper both analyzes the differences in Web sources that are cited by generative and traditional search engines, and further examines each engine's internal consistency and robustness to small variations in query syntax. Furthermore, our analysis is conducted on a larger scale (with 14,212 queries in total) than any existing work. The goal of our study is to provide insights for website publishers regarding potential responses to generative AI, to help users better understand the differences and trade-offs between traditional and generative search, and to call for collaboration toward  building a sustainable online publishing and generative AI ecosystem.

\section{Empirical Setup}
\label{sec:experimental_setup}
%
%
\yc{DONE: if have time, change to Amazon Retail-Q or Amazon Retail:Q, same for Comp}
\yc{Done: below, when you say "sampled", better specify if its random sample or any criteria used in the sampling}
\textbf{Benchmark Query Dataset}. Our benchmark contains 11{,}500 queries, which expands upon existing query datasets used to study generative search engines~\cite{puerto2025cseo,aggarwal_geo_2024} to specifically investigate the effects of query syntax (e.g., keywords vs. natural language question) and intent (e.g., asking for product information rather than where to buy). We further add coverage of localized queries (e.g., ``near me'' queries). As summarized in Table~\ref{tab:dataset_components}, we partition the benchmark into 9 query categories: \textbf{(1) ORCAS}: representative real-user queries~\cite{craswell2020orcas} labeled with query intent (i.e., informational, navigational, or transactional)~\cite{alexander_few_2025}; \textbf{(2) Amazon Retail}: real-user Amazon Retail keyword queries~\cite{puerto2025cseo,reddy2022shopping}; \textbf{(3) Amazon Retail-Comp}: product-comparison queries based on Amazon Retail queries; \textbf{(4) Amazon Retail-Q}: product questions based on Amazon Retail queries; \textbf{(5) Debate}: debate-style queries~\cite{liu-etal-2023-evaluating}; \textbf{(6) ELI5}: complex informational queries randomly sampled from a Reddit-based corpus~\cite{fan2019eli5longformquestion}; \textbf{(7) Localized}: randomly sampled from ORCAS ``near me'' queries~\cite{craswell2020orcas} and instantiated with representative cities; \textbf{(8) NQ}: randomly sampled Natural Questions \cite{aggarwal_geo_2024}; and \textbf{(9) NQ Keywords}: keyword-style reformulations of NQ queries. We used Gemini~2.5 Flash to generate queries for categories 3, 4 and 9.\footnote{Our repository contains all queries, as well as prompting templates and settings used to generate synthetic query variants.} Two authors manually viewed 50 of the generated queries to ensure that Gemini produced outputs that changed the query structure while preserving the original query's topic.

\newcommand{\sample}[1]{{\scriptsize \emph{#1}}}

\begin{table}[!t]
\centering
\footnotesize
\setlength{\tabcolsep}{1pt}
\renewcommand{\arraystretch}{1}
\caption{Composition of the Benchmark Query Set}
\label{tab:dataset_components}
\begin{tabular}{l c p{5cm}}
\toprule
\textbf{Dataset} & \textbf{\# Queries} & \textbf{Sample} \\
\midrule
ORCAS & 5{,}000 & \sample{public aid office locations} \\
Amazon Retail & 500 & \sample{star lamp projector galaxy} \\
Amazon Retail-Comp & 500 & \sample{Compare star lamp projector galaxy vs aurora projector} \\
Amazon Retail-Q & 500 & \sample{What is the best star lamp projector for a small room?} \\
Debate & 1{,}000 & \sample{Should LGBT rights be protected by law?} \\
ELI5 & 1{,}000 & \sample{What determines if a phosphorylated protein is on or off?} \\
Localized & 1{,}000 & \sample{free rabies shot near Alicante, Spain} \\
NQ & 1{,}000 & \sample{when was the first cell phone call made} \\
NQ Keywords & 1{,}000 & \sample{first cell phone call} \\
\midrule
\textbf{Total} & \textbf{11{,}500} & \\
\bottomrule
\end{tabular}
\end{table}

\textbf{Collecting SERP and AIO Responses.}
To enable a controlled, large-scale comparison between traditional SERP and AIO, we follow prior work \cite{press-etal-2023-measuring,yoran-etal-2023-answering,zhao-etal-2025-uncertainty} and use a search retriever (i.e., SerpAPI) to collect representative results for real-users (see Section~\ref{serpapi_robustness} for its robustness test). For each query in our benchmark query set, we issue a Google Search request via SerpAPI to collect the SERP and AIO (when available) results. The retriever is instructed to simulate a mobile device from Newark, NJ to ensure that SERP and AIO sources are collected under identical timing, device, and localization conditions. We confirm that our findings are robust to changes in device and location in Section~\ref{sec:robustness}. To reflect real world-importance, we only consider the SERP results from the first page because over 97\% of all clicks are made on the first page of search outputs~\cite{urman2021andwheresearchcomparative}.

\textbf{Collecting Gemini Responses.} 
Using the Gemini API, we collect AI chatbot responses from the Gemini 2.5 Flash model. We selected this model because it provides a good trade-off between price and performance, and AIOs are built with a similar lightweight Gemini model. Grounding with Google Search is enabled so that the model will retrieve sources for each query. In alignment with AIO, the Thinking mode, typically used for complex reasoning tasks, is disabled. The input text consists of the given query from our dataset only, with no further prompting; no custom system instructions are added. Finally, we adhere to the default inference parameters. In the interest of comparability, all SERP, AIO, and Gemini responses to benchmark queries are collected on December 7th-8th 2025. 


\textbf{Evaluation Metrics.}
\label{sec:metrics}
 To quantify discrepancies in the retrieved sources between two search engines, we compute both set-level and rank-sensitive overlap. \textbf{Jaccard similarity} compares the set of unique sources, ranging from 0 (no shared sources) to 1 (identical sets), and captures whether the same sources appear regardless of rank. 
\textbf{Rank-biased overlap (RBO)} \cite{Webber2010ASM} compares ranked lists (also ranging from 0 to 1) while placing more weight on agreement near the top of the ranking, reflecting that higher-ranked sources tend to receive more user attention. We compute both metrics at the URL-level. Both Jaccard similarity and RBO were chosen in part because they are naturally capable of handling lists of differing length. Following the recommended implementation of RBO for public search engines that output around 10 results~\cite{Webber2010ASM}, we set the persistence parameter to $p=0.9$.

\section{Empirical Analysis}

\begin{figure*}[t!]
\includegraphics[width=\textwidth]{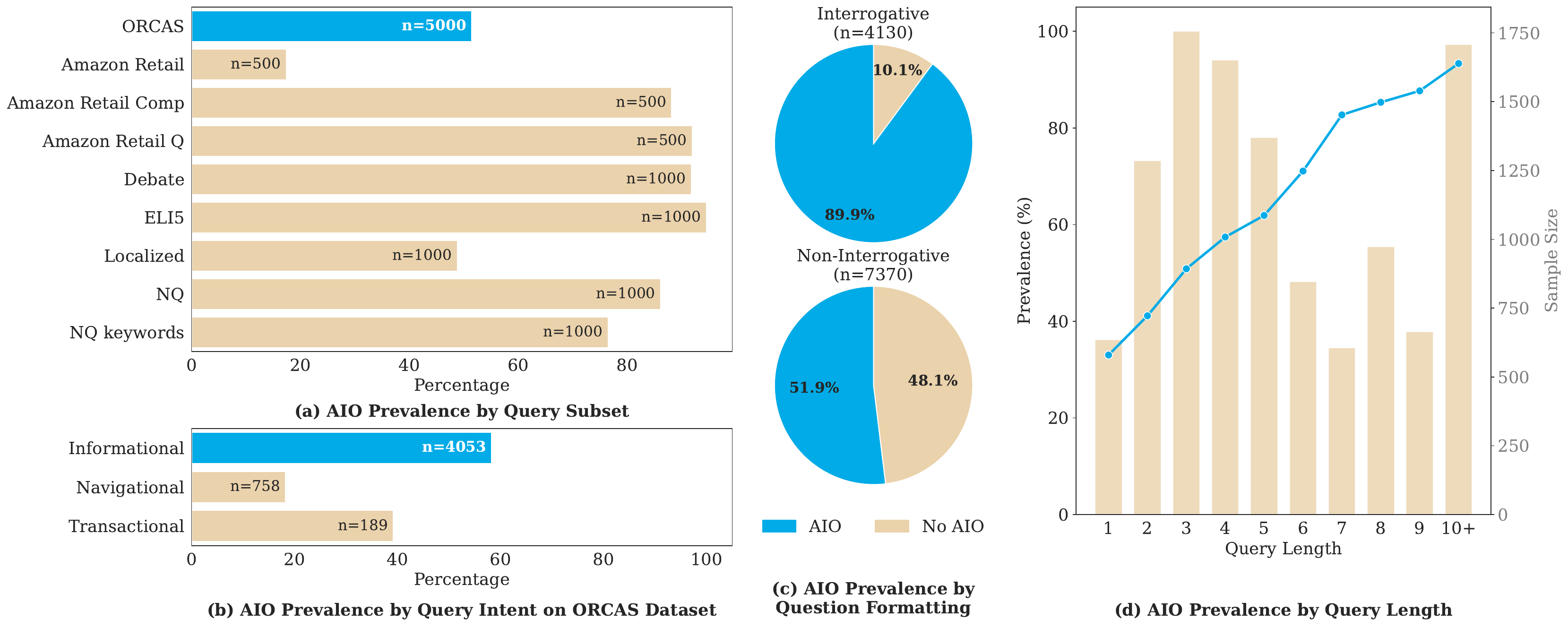}
    \caption{\centering AIO Prevalence and Factors Associated with Higher Likelihood of an AIO Being Generated \newline Note: In panel (d), the blue line represents AIO prevalence and the bar chart shows the number of queries per length.}
    \label{fig:AIO_prevalence}
\end{figure*}

This paper aims for a better understanding of how generative AI is disrupting Web search and the resulting implications for websites and users. Toward this goal, our empirical analysis is guided by these research questions:

\begin{itemize}
    
    \item[RQ1:] How frequently are AIOs generated, and for what types of queries?
    \item[RQ2:] How similar are the sources retrieved by AIO, SERP, and Gemini?
    \item[RQ3:] How do the characteristics of retrieved sources differ between generative and traditional search engines?
    \item[RQ4:] How consistent and robust are generative search engines, relative to traditional search?
    \item[RQ5:] How does generative search handle high-stakes queries?

\end{itemize}

\subsection{RQ1: AIO Generation}
\label{subsec:aio_generation}

We find AIOs to be generated on 65.6\% of all benchmark queries. In Figure~\ref{fig:AIO_prevalence}a, we show that the AIO generation rate is substantially different across the different query subsets in our benchmark dataset.  On the extremes, 94.6\% of ELI5 queries, compared to just 17.4\% of Amazon Retail queries, result in an AIO. We consider the 51.5\% AIO generation rate on the ORCAS dataset to be the most representative of how often real-user queries result in a generated AIO. 



We also show that the AIO generation rate varies based on query characteristics such as intent, format, and length. 
In Figure~\ref{fig:AIO_prevalence}b, we utilize the assigned query intent categories from the ORCAS dataset and find that informational queries are most likely to lead to an AIO. The AIO generation rates are significantly different across categories ($\chi^2(2,N=5000)=439.56,p<0.001$), and all post-hoc pairwise comparisons with the Bonferroni correction were significant (p<0.001). 

In Figure~\ref{fig:AIO_prevalence}c, we separate out interrogative queries (i.e., those that end with a question mark or contain interrogative words such as who, what, or when) from non-interrogative queries. Interrogative queries are significantly more likely to result in an AIO being generated ($\chi^2(1,N=11500)=1685.7,p<0.001$). This trend may also be found in the AIO generation rates of NQ (86.2\%) and NQ keywords (76.5\%) queries, as the only difference is NQ queries are formatted as a question, and NQ Keywords queries are formatted as a list of the keywords derived from the question.

Figure~\ref{fig:AIO_prevalence}d shows that across the full benchmark dataset, the AIO prevalence increases with query length. We confirm that this trend also exists within the ORCAS queries to ensure it is not simply a result of differing average lengths for each query subset.

Taken together, these findings show that AIOs are more likely to be generated when users seek information, and particularly for queries that specify exactly what the user is looking for (e.g., ``How do I clean a pizza stone?'') instead of broad informational queries (e.g., ``pizza stone'').   


For the rest of our analyses, we focus on the 7,439 queries where Google Search, AIO, and Gemini returned sources. We excluded the 56 and 121 queries where the AIO or Gemini response was generated without sources, respectively. This included a mix of queries where internal knowledge was relied on (e.g., ``show me all the letters in the English alphabet) and queries without meaningful responses (e.g., Gemini cannot answer the query ``lunch spots near me'' because it cannot access user location). 

\subsection{RQ2: Similarity Between Traditional and Generative Search Engine Sources}
\label{subsec:rbo_jaccard_ndcg}
\begin{table}[!t]
\centering
\begin{tabular}{l c c c c c c}
\hline
\multirow{3}{*}{\textbf{Query Subset}} & \multicolumn{3}{c}{\textbf{Jaccard}} & \multicolumn{3}{c}{\textbf{RBO}}\\
\cmidrule(lr){2-4} \cmidrule(lr){5-7} 
 & AIO & AIO & GEM & AIO & AIO & GEM  \\
 & SERP & GEM & SERP & SERP & GEM & SERP \\
\hline
ORCAS         & \cellcolor{red!51}0.17 &  \cellcolor{red!36}0.12 & \cellcolor{red!60}0.20 & \cellcolor{red!72}0.24  & \cellcolor{red!51}0.17  & \cellcolor{red!83}0.26 \\
Amazon Retail          & \cellcolor{red!19}0.08 &  \cellcolor{red!6}0.06 & \cellcolor{red!19}0.08 & \cellcolor{red!36}0.12  & \cellcolor{red!30}0.10  & \cellcolor{red!30}0.10 \\
 Retail-Comp         & \cellcolor{red!33}0.11 &  \cellcolor{red!19}0.08 & \cellcolor{red!19}0.08 & \cellcolor{red!45}0.15  & \cellcolor{red!30}0.10  & \cellcolor{red!30}0.10  \\
 Retail-Q    & \cellcolor{red!36}0.12 &  \cellcolor{red!30}0.10 & \cellcolor{red!30}0.10 & \cellcolor{red!47}0.16  & \cellcolor{red!42}0.14  & \cellcolor{red!39}0.13 \\
Debate       & \cellcolor{red!77}0.24 &  \cellcolor{red!30}0.10 & \cellcolor{red!33}0.11 & \cellcolor{red!88}0.27  & \cellcolor{red!36}0.12  & \cellcolor{red!42}0.14  \\

ELI5   & \cellcolor{red!63}0.21 &  \cellcolor{red!33}0.11 & \cellcolor{red!42}0.14 & \cellcolor{red!82}0.25  & \cellcolor{red!45}0.15  & \cellcolor{red!54}0.18 \\

Localized    & \cellcolor{red!42}0.14 &  \cellcolor{red!15}0.07 & \cellcolor{red!48}0.16 & \cellcolor{red!51}0.17  & \cellcolor{red!30}0.10  & \cellcolor{red!68}0.21 \\

NQ        & \cellcolor{red!57}0.19 &  \cellcolor{red!39}0.13 & \cellcolor{red!57}0.19 & \cellcolor{red!72}0.24  & \cellcolor{red!51}0.17  & \cellcolor{red!83}0.26 \\

NQ Keywords        & \cellcolor{red!57}0.19 &  \cellcolor{red!39}0.13 & \cellcolor{red!60}0.20 & \cellcolor{red!72}0.24  & \cellcolor{red!48}0.16  & \cellcolor{red!83}0.26 \\

\hline
\hline
\textbf{Total}          & \cellcolor{red!54}\textbf{0.18}    & \cellcolor{red!33}\textbf{0.11}  & \cellcolor{red!48}\textbf{0.16} & \cellcolor{red!69}\textbf{0.23} & \cellcolor{red!45}\textbf{0.15}  & \cellcolor{red!64}\textbf{0.21} \\
\hline
\end{tabular}
\caption{\centering Average Similarity in Returned Sources for Google AI Overviews (AIO), Traditional Google Search Results (SERP), and Gemini 2.5 Flash (GEM) }
\label{tab:dataset_rbos}
\end{table}
Table~\ref{tab:dataset_rbos} presents the average similarity between the list of sources returned by the AIO, Gemini, and traditional SERP for each query in the benchmark dataset. The main takeaway is that \emph{regardless of query subset and which pair of search engines is compared, the retrieved lists are dissimilar, despite all three being developed by Google.}
%
Besides using RBO as a metric for comparing the similarities of ranked lists, we also present Jaccard similarity because of its interpretability. 
For example, our findings show that on average, only 18\% of the sources returned by either the AIO or traditional SERP will be retrieved by both search engines. Similar trends are observed when source similarity is measured by RBO. Surprisingly, despite AIO being built with a lightweight Gemini model, the source lists retrieved by AIO and Gemini are the least similar.

We also evaluate the retrieval similarity for different query subsets. Although the similarity metric values vary by query subset, the amount of similarity remains low. For example, no search engine pairing has an average RBO greater than 0.27 (i.e., AIO and SERP for Debate queries). In particular, the product and service queries (i.e., Amazon Retail and Localized query subsets) result in the lowest similarity for each pairing of search engines. 


To contextualize the dissimilarity, we report the average number of sources returned by each  search engine: 9.68 for Gemini, 9.24 for AIO, and 8.75 for traditional SERP. As these numbers are similar, we attribute the low similarity between source lists to different methodologies for each search engine rather than a different number of retrieved sources. Furthermore, the inconsistency between search engines greatly exceeds the inconsistency between two runs of the same search engine (see Section~\ref{subsec:consistency}), so we cannot attribute this dissimilarity to built-in randomness. 
%

%
%
\begin{figure}[t!]
\includegraphics[width=\linewidth]{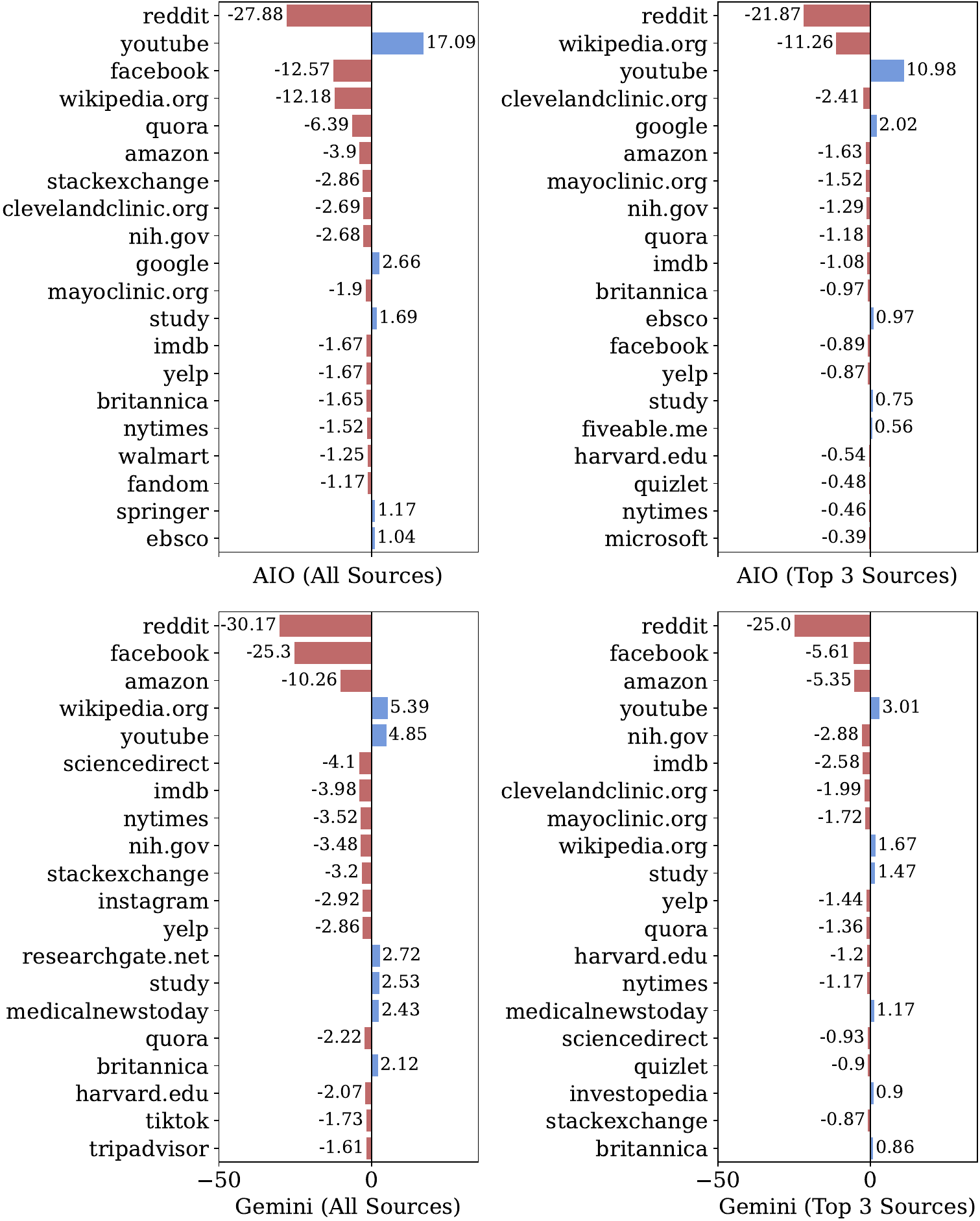}
    \caption{\centering Domains with Biggest Change in Prevalence (\% of Queries Retrieved) Relative to Traditional Search}
    \label{fig:domain_prevalence_2_x_2}
\end{figure}
\subsection{RQ3: Characteristics of Generative and Traditional Search Sources}
\label{subsec:source_comparisons}

After showing the large discrepancy in the sources retrieved by AIOs, Gemini and traditional SERP in Section~\ref{subsec:rbo_jaccard_ndcg}, we further study  which websites, or categories of websites, benefit from this disruption. In particular, we determine how the popularity, type of content, and website's decision on blocking Google's AI bot are associated with generative search engine sources in comparison to traditional search sources. 

%
\subsubsection{Most Affected Domains}

In Figure~\ref{fig:domain_prevalence_2_x_2}, we present the domains with the largest absolute changes in the percentage of benchmark queries for which they are retrieved by generative search engines compared with traditional search engines. The subfigures on the right side of Figure~\ref{fig:domain_prevalence_2_x_2} present the changes when considering only the top three ranked sources, rather than all sources, retrieved by each search engine. We consider the top \emph{three} specifically because Google AIO shows three sources automatically. Additional sources beyond the top three can only be accessible when users  click the ``Show all'' button or specific link icons (See Figure~\ref{fig:aio_interface}). Furthermore, ranking has been shown to significantly affect the click-through-rate of traditional search engine results~\cite{zilincan2015search,baye2016search,seo_ieee_paper}, and prior research and SEO professionals suggest this applies to generative search as well~\cite{aggarwal_geo_2024,aio_top3,marketing_couch}. 

We have three main takeaways from Figure~\ref{fig:domain_prevalence_2_x_2}. First, large and well-known websites are the most affected (both positively and negatively). This is intuitive as large websites have the reputation and diversity in content to be relevant to many different queries. Second, the overwhelming majority of these websites receive fewer overall, and fewer top three, citations with generative search engines (indicated by red bars and negative numbers in Figure~\ref{fig:domain_prevalence_2_x_2}). This suggests that generative search tends to source information from more niche sources than traditional search engines. Third, Google's AIOs favor Google websites (i.e., google.com and youtube.com domains). Gemini also favors YouTube in comparison to traditional Google Search, but the absolute difference is smaller. Lastly, we note that McNemar's tests~\cite{McNemar_1947} confirm that all differences in Figure~\ref{fig:domain_prevalence_2_x_2} are statistically significant. 
%

%




%
\subsubsection{Blocking the Google-Extended Bot}
In our analyses of the most affected domains, we found that 21 popular publishers listed in Table~\ref{tab:robots_domains} (which are retrieved for at least 20 unique queries by both Google Search and AIOs) were never cited by Gemini. Several popular social media (Facebook, Instagram, Tiktok) and review websites (IMDb, Yelp, Tripadvisor) also received zero citations from Gemini. Upon further investigation, we found that all of these websites block the Google-Extended bot in their robots.txt files. 

Google states that websites can block this bot if they do not want their content to be used for training future Gemini models, nor for grounding responses with sources (i.e., citations). Google further clarifies that this will not affect a websites' ranking in Google Search results, nor does it impact the AIO's access to this content~\cite{google_crawlers}. Thus, our results indicate that Google does respect websites' wishes, and this means that the decline in visibility in Gemini search results is self-inflicted. However, a surprising finding is that many of these websites also have fewer citations in AIOs (e.g., NYTimes, Yelp, IMDb, and Facebook). 
%
%
\begin{table}[t]
    \centering
    \small
    \begin{tabular}{|c|c|c|}
    \hline
         NYTimes & ESPN & Genius \\
        CNN & Business Insider & National Geographic \\
        BBC & CNBC & The Conversation \\
        ScienceDirect & NPR & U.S. News \& World Report \\
        Reuters & WIRED & Scientific American \\
        Wiley & USA Today & Consumer Reports \\
       Nature & NBC News & STAT \\
        %
    \hline
    \end{tabular}
    \caption{\centering Popular Publishers Not Used by Gemini (Ranked by Tranco Popularity)}
    \label{tab:robots_domains}
\end{table}
\subsubsection{Does Generative Search Favor Niche Websites?}

From Figure~\ref{fig:domain_prevalence_2_x_2} we see that AIOs and Gemini tend to rely less on many of the popular domains that are retrieved by traditional Google Search for a wide range of queries. We are interested in understanding whether AIO and Gemini reference more niche websites or just a different set of popular websites. 

\begin{figure}[t!]
\includegraphics[width=\linewidth]{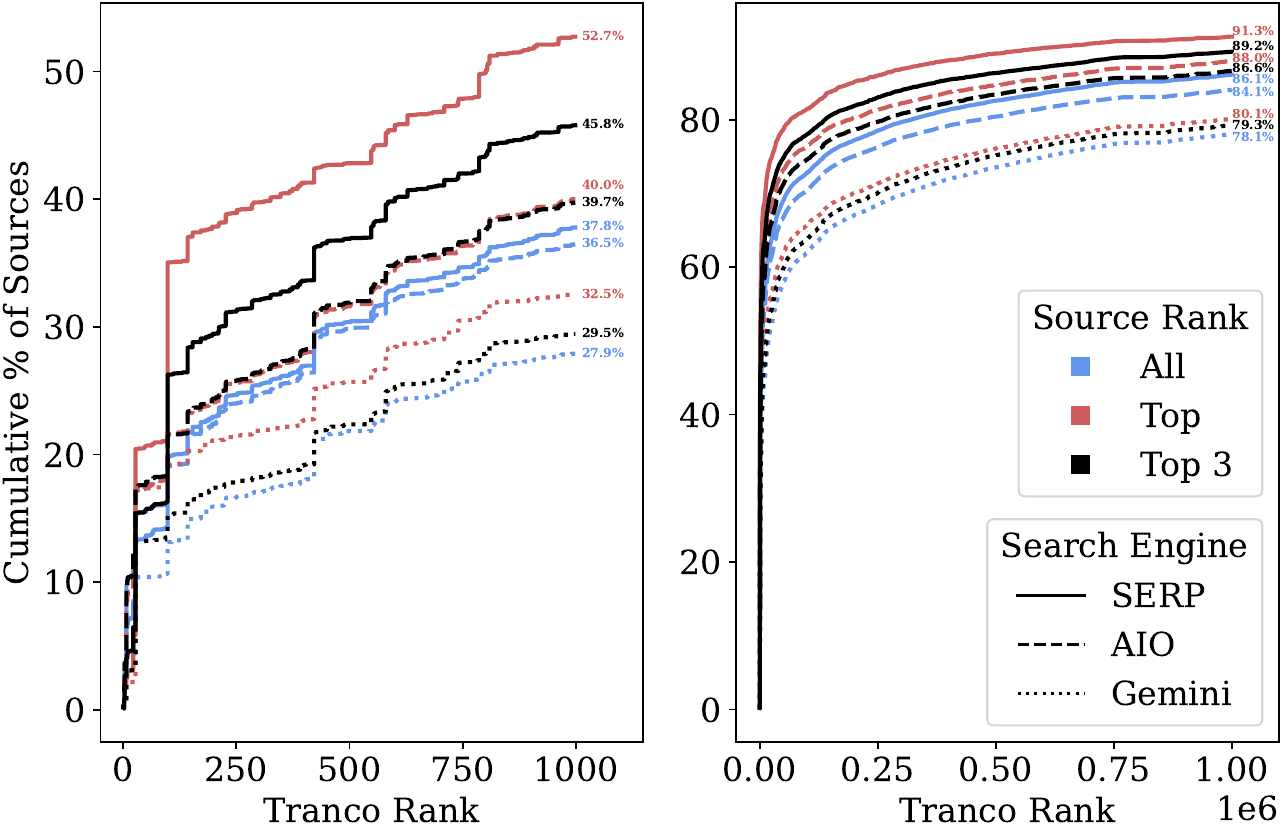}
    \caption{\centering Percentage of Retrieved Sources From the Top 1K (left) and 1M (right) Popular Domains}
    \label{fig:tranco_cdfs}
\end{figure}

To analyze this, we use the Tranco rankings~\cite{tranco}, which are designed to be an objective ranking of website popularity for academic research\footnote{We use the list from December 7th, 2025: \url{https://tranco-list.eu/list/6GWVX/1000000}}. In Figure~\ref{fig:tranco_cdfs}, we plot each search engines' cumulative percentage of retrieved sources across all benchmark queries as a function of the domains' Tranco rank.

Figure~\ref{fig:tranco_cdfs} (left) focuses on the top 1,000 domains in terms of popularity. Google Search retrieves 37.8\% of all sources from the top 1,000, which is 1.3 and 9.9 percentage points higher than AIO and Gemini, respectively. This gap substantially increases when considering just the top one or top three retrieved sources from each search engine. 
Considering only the top ranked source for each query, websites ranked in the top 1,000 are cited in response to 52.7\% of queries for traditional SERP, and just 40.0\% and 32.6\% for AIO and Gemini, respectively. 
This suggests that traditional search places higher importance on source reputability and popularity.

\begin{figure}[t!]
\includegraphics[width=\linewidth]{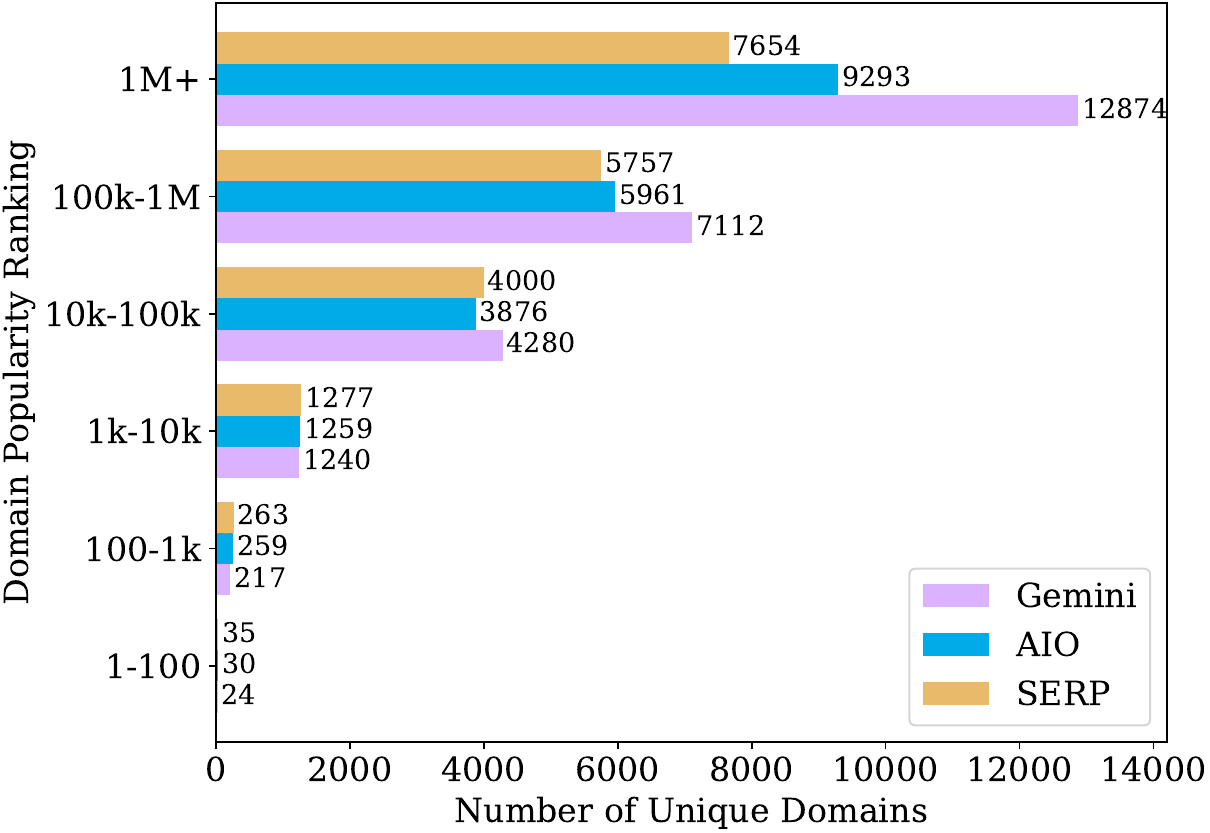}
    \caption{Number of Unique Sources by Tranco Rank}
    \label{fig:domains_by_tranco}
\end{figure}

Figure~\ref{fig:tranco_cdfs} (right) zooms out to consider the top 1 million domains in terms of popularity. The overall trend of traditional search retrieving a higher percentage of popular domains is consistent. However, the gaps have narrowed, particularly when looking at only the top one or top three retrieved sources. Figure~\ref{fig:domains_by_tranco} plots the number of unique domains referenced by each search engine binned by Tranco rank. Despite having fewer overall unique sources, we see that traditional Google Search has the highest number of unique sources ranked in the top 100, top 1K, and top 10K of Tranco. Gemini has the most unique sources for domains ranked outside the top 10K. The AIO has more unique sources than traditional Google Search for domains ranked outside the top 100K. 
%

    
%
%
\subsubsection{Statistical Significance of Differences}

\begin{figure}[t!]
\includegraphics[width=\linewidth]{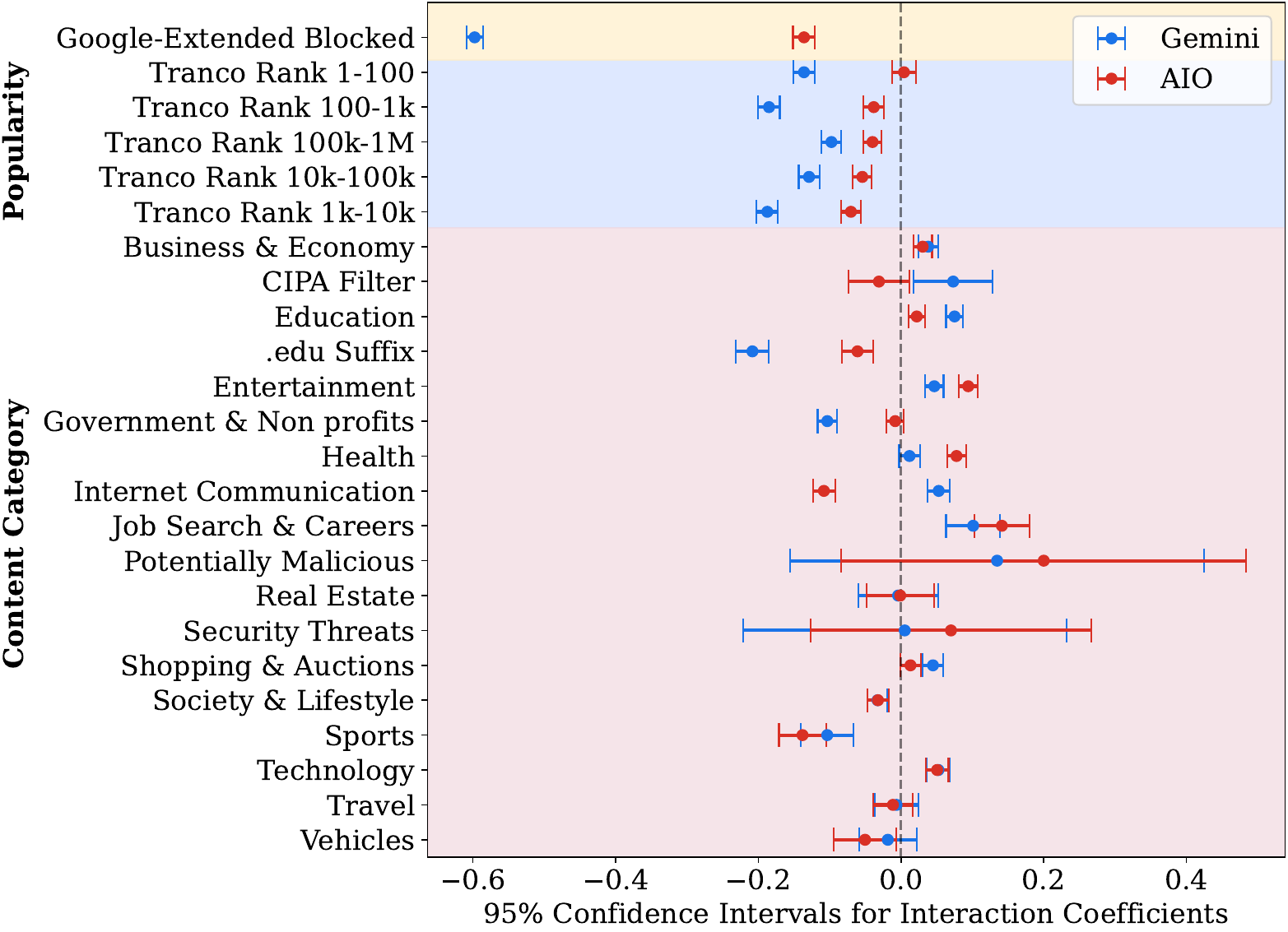}
    \caption{\centering Characteristics of Generative Search Sources Relative to Traditional Search Sources}
    \label{fig:regression}
\end{figure}

We test for statistically significant differences in the website characteristics of sources retrieved by generative versus traditional search engines. We construct a dataset with every unique pair of source and query, and mark whether each of the three search engines cited the source for the particular query. 
Importantly, because all relevant sources had to be retrieved by at least one search engine to be included in the dataset, the results can only be interpreted in comparison to the other search engines. We compare each generative search engine to the traditional search engine using a linear probability model with query fixed effects and clustered standard errors. We regress whether a source is cited for a query onto the interaction variables between search engine type and website characteristics. In particular, we include the binned tranco rank, whether the website blocks the Google-Extended bot, and 17 Cloudflare domain categories (see Figure~\ref{fig:regression}) that are included in the list of retrieved sources for at least 5\% of all queries. We also add whether the domain ends in .edu because the ``Education'' category in Cloudflare includes encyclopedia and Q\&A websites, which may be seen as less reputable than content produced by educational institutions.

The 95\% confidence intervals for all website characteristics interacted with the two generative search engines are displayed in Figure~\ref{fig:regression}. Coefficients are statistically significant if the corresponding bar does not cross zero. To summarize, the first takeaway is that both Gemini and AIOs are significantly less likely to retrieve content from websites blocking the Google-Extended crawler, despite AIOs technically having access to this content. Second, we find that Gemini and AIO are both significantly less likely to retrieve content from popular websites. Interestingly, this difference is most substantial among domains with a Tranco ranking from 1k-10k, which corresponds to websites that receive substantial traffic, but are not necessarily well-known. Third, although generative search is significantly less likely to cite reputable institutions in government or education, it is not significantly more likely to source content from inherently bad websites (e.g., those with security threats). However, Gemini is significantly more likely to cite content from websites that are inaccesible to children in publicly funded facilities via the Children's Internet Protection Act (CIPA). 

\subsection{RQ4: Search Engine Inconsistency}
\label{subsec:consistency}
Besides comparing the sources retrieved by generative and traditional search engines, we also evaluate the consistency of the sources retrieved between two runs, particularly when queries are run from a different type of device or after minor query edits. High inconsistency between outputs in response to queries that have the same intent is problematic at a societal-level as different users receive different information to the same questions. It can also be an undesirable behavior at an individual-level as it can make it difficult to find previously retrieved information again.

%
\begin{table}[t]
\centering
\begin{tabular}{l c c c c c c}
\hline
\multirow{2}{*}{\textbf{Device/}} & \multicolumn{3}{c}{\textbf{Jaccard}} & \multicolumn{3}{c}{\textbf{RBO}}  \\
\cmidrule(lr){2-4} \cmidrule(lr){5-7} 
 \textbf{Location} & AIO & GEM & SERP & AIO & GEM & SERP \\
\hline
Both Same          & \cellcolor{red!46}0.66 & \cellcolor{red!12}0.46 & \cellcolor{red!71}0.78 & \cellcolor{red!53}0.69 & 
\cellcolor{red!21}0.52 & \cellcolor{red!95}0.86 \\
Diff Location      & \cellcolor{red!44}0.64 &  - & \cellcolor{red!68}0.76 & \cellcolor{red!52}0.68 & - &\cellcolor{red!88}0.84  \\
Diff Device   & \cellcolor{red!25}0.55 &  - & \cellcolor{red!53}0.69 & \cellcolor{red!23}0.53 & - & \cellcolor{red!72}0.79 \\
Both Diff       & \cellcolor{red!25}0.55 & - &  \cellcolor{red!47}0.67 & \cellcolor{red!23}0.53 & - & \cellcolor{red!70}0.77   \\
\hline
\end{tabular}
\caption{\centering Average Similarity in Returned Sources Between Runs and with Different Device or Location}
\label{tab:rbo_consistency}
\end{table}
\textbf{Consistency Across Runs, Devices, and Locations}. We first selected a stratified random sample of 100 queries from our benchmark dataset. We then simultaneously collected AIO and traditional search responses two times for each configuration of device (mobile and desktop) and location (Austin, TX and Newark, NJ). Since Gemini does not consider device or location, we just run the same query twice. We display the average similarity between the sources retrieved by each search engine across runs, devices, and locations, in terms of Jaccard similarity and RBO, in Table~\ref{tab:rbo_consistency}. In all possible comparisons from Table~\ref{tab:rbo_consistency}, the generative search engines show lower consistency than traditional Google Search. 

\yc{give the numbers in the above sentence. for the next sentence, I don't get the point of "different user interfaces" playing a role and suggest to remove}
Table~\ref{tab:rbo_consistency} also shows that for traditional SERP and AIOs, the retrieved sources are more inconsistent between different devices than between different locations (which only decreases similarity metrics by 0.01-0.02). The large effect of device type may be due to Google's expectation that mobile users seek different information from desktop users. We further find that changing device results in more inconsistency for AIOs than traditional SERP (i.e., relative to RBO for the same device and location, RBO decreases by 0.16 and 0.07 for AIO and SERP, respectively, when changing device). This is largely driven by the finding that AIOs retrieve more sources on average for mobile searches (10.38) than desktop searches (9.42), while traditional Google Search retrieves similar amounts for both (8.79 vs. 8.65). This is counterintuitive, as we would expect fewer AIO sources on mobile due to the smaller screen size.

\yc{I don't get the point of the preceding sentence. AIO is more useful for mobile users than desktop users does not lead to more sources retrieved for mobile search than desktop search }
%
\yc{Add a few sentence about location impact. }
%
\yc{the paragraph below. the contraction is written as each way, but abbreviation and question mark are written as two way. is that intentional? if not, I think it's easier to write as two way for all those cases}
\yc{SerpAPI vs SERP, use one form, consistently}
\textbf{Consistency in the Presence of Cosmetic Query Edits}. \label{cosmetic_changes} We first make minor edits to 200 queries: 100 queries where two words are (un)contracted (e.g., ``what is'' vs ``what's''); 50 queries where a word is (un)abbreviated (e.g., ``United States'' vs ``U.S.''); and 50 queries where a question mark is added (removed) to queries that are clearly questions regardless of punctuation. Our findings reveal that AIOs are less robust to these changes. On average, the retrieved sources from the original and modified query have an RBO of 0.49 for AIOs (a 28.99\% decline in comparison to the RBO resulting from two runs of the same query). In contrast, traditional SERP and Gemini have average RBOs of 0.74 and 0.5, respectively (13.95\% and 3.85\% declines, respectively). One potential explanation for why this issue plagues AIOs, and not Gemini, is that AIOs utilize a lightweight Gemini model with lower reasoning capabilities, which may result in more reliance on query keywords than intent.

We further analyze the impact of inconsistencies in retrieved sources to the text generated by generative search engines. We quantify this relationship using word-level Jaccard similarity.\footnote{Each text is lowercased, tokenized with NLTK's \texttt{wordpunct\_tokenize}, and Porter-stemmed (\texttt{PorterStemmer}).} We find a strong and statistically significant positive Pearson correlation between source similarity and generated summary similarity. For AIOs, higher RBO in the retrieved sources between original and edited queries corresponds to higher text similarity (i.e., $r=0.62$, $p<0.001$). Gemini exhibits the same pattern (i.e., $r=0.55$, $p<0.001$). In summary, even when the query’s underlying intent is unchanged, cosmetic query differences result in different sources retrieved, and subsequently different generated text.

%
%
\begin{figure}[t!]
\includegraphics[width=\linewidth]{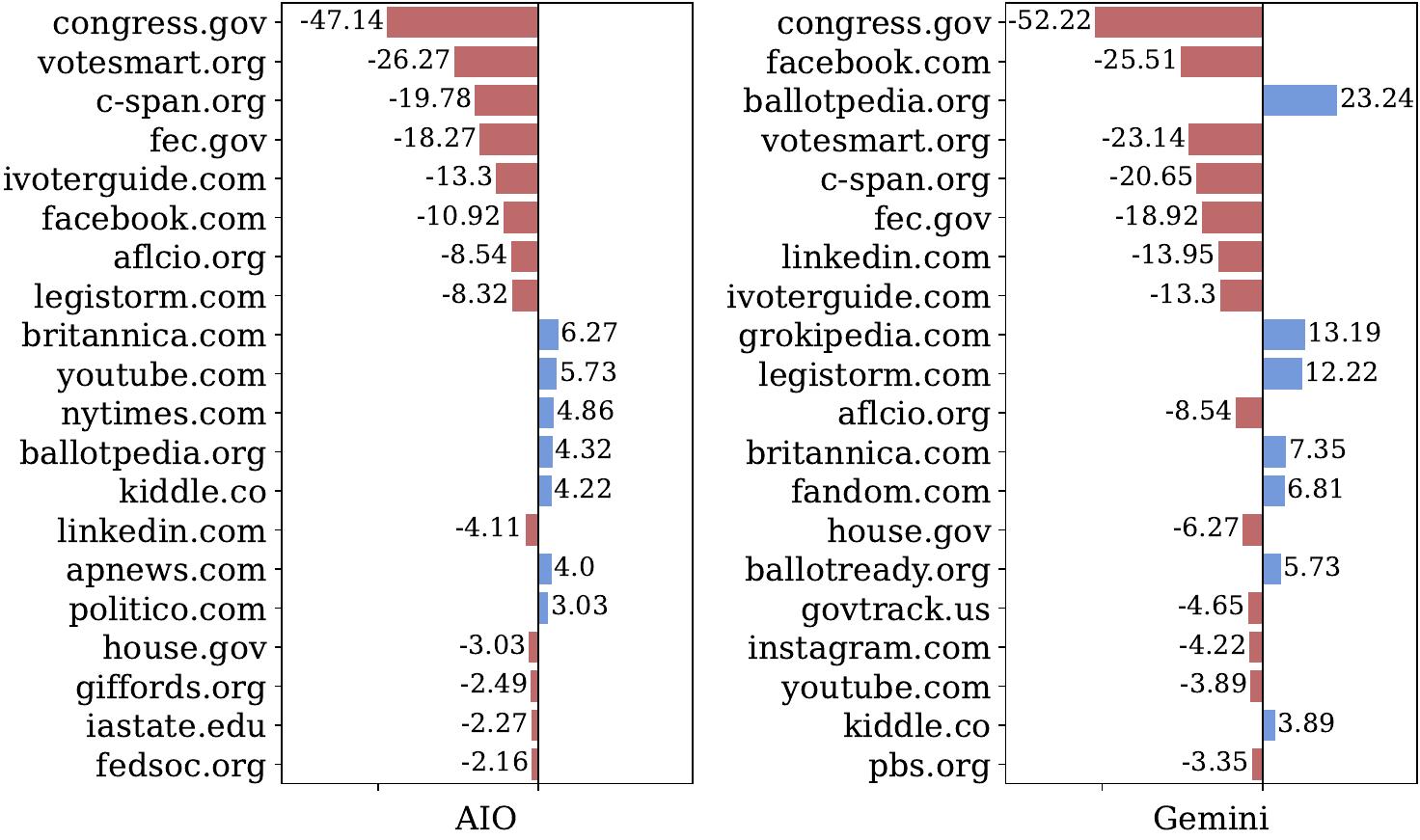}
    \caption{\centering Domains with Biggest Change in Prevalence (\% of Queries Retrieved) for Political Queries}
    \label{fig:pol_queries}
\end{figure}
\subsection{RQ5: High-Stakes Queries}
\label{subsec:politicians_and_trends}

To further understand how generative search affects users, we evaluate how generative search responds to high-stakes queries where errors could have societal impact. Specifically, we consider queries related to debate topics and political figures due to their sensitive nature and potential for different viewpoints. We also add trending queries, which are particularly susceptible to misinformation due to the lack of consensus as events unfold in real time. 

While the debate queries are part of our benchmark dataset, political and trending queries were not included in the benchmark dataset due to their time-sensitive nature. For political queries, we generate queries of the form ``What is the politician [NAME] known for?'' for all current members of the U.S. Congress, as well as those who finished as runner-up in the last election. We curated this new list of queries to differentiate it from the debate query set, and other similar political queries in prior research~\cite{kirsten2025,dai-etal-2025-media,steiner2022seek}, which focus on political ideas or events, rather than politicians. For trending queries, we utilized 863 trending topics from Google Trends\footnote{\url{https://trends.google.com/trending}} with at least 1,000 searches and 2+ subqueries (collected December 19th, 2025). For each trending topic, we collected data for the main query, and one randomly selected subquery. 

\textbf{Debate Queries}. As shown in Figure~\ref{fig:AIO_prevalence}a, AIOs are frequently generated for these controversial topics. Surprisingly, a substantial percentage (33.4\%) of the AIO summaries started with affirmative or negative responses. This was less common, but not absent, in Gemini responses (5.6\%). Even if both sides of the debate were presented further down in the summary, it is surprising to see Gemini or AIO taking a stance in response to controversial questions such as (i) whether robots should be included in the military, (ii) whether AI should be used for medical diagnoses, (iii) whether meat consumption should be restricted to reduce global warming, and (iv) whether immigration laws should be reformed.


    

\textbf{Political Queries}. AIOs are generated frequently for politicians (93.8\%). Concerningly, generative search engines may retrieve less credible sources for this crucial information. 
Figure~\ref{fig:pol_queries} presents the domains exhibiting the largest deviations from those retrieved by traditional search for political queries.
As shown in Figure~\ref{fig:pol_queries}, the generative search engines cite government resources (e.g., congress.gov) less frequently. The AIO retrieves more content from popular news domains like NYTimes, Politico, and AP News, while Gemini has larger reliance on Fandom and Grokipedia.

We further attempt to quantify this issue using News source credibility ratings from Media Bias/Fact Check (MBFC)\footnote{\url{https://mediabiasfactcheck.com/methodology/}}. Although they only provide ratings for about half of all sources retrieved for political queries, we still find that more sources retrieved by the generative search engines have questionable (i.e., medium or low) ratings. Specifically, 11.4\% (0.5\% low credibility) of sources retrieved by AIO and 15.0\% (0.6\% low) of sources retrieved by Gemini come from websites with questionable credibility ratings, in comparison to 10.6\% (0.2\% low credibility) of sources retrieved by traditional search. 

\begin{figure}[t!]
\includegraphics[width=\linewidth]{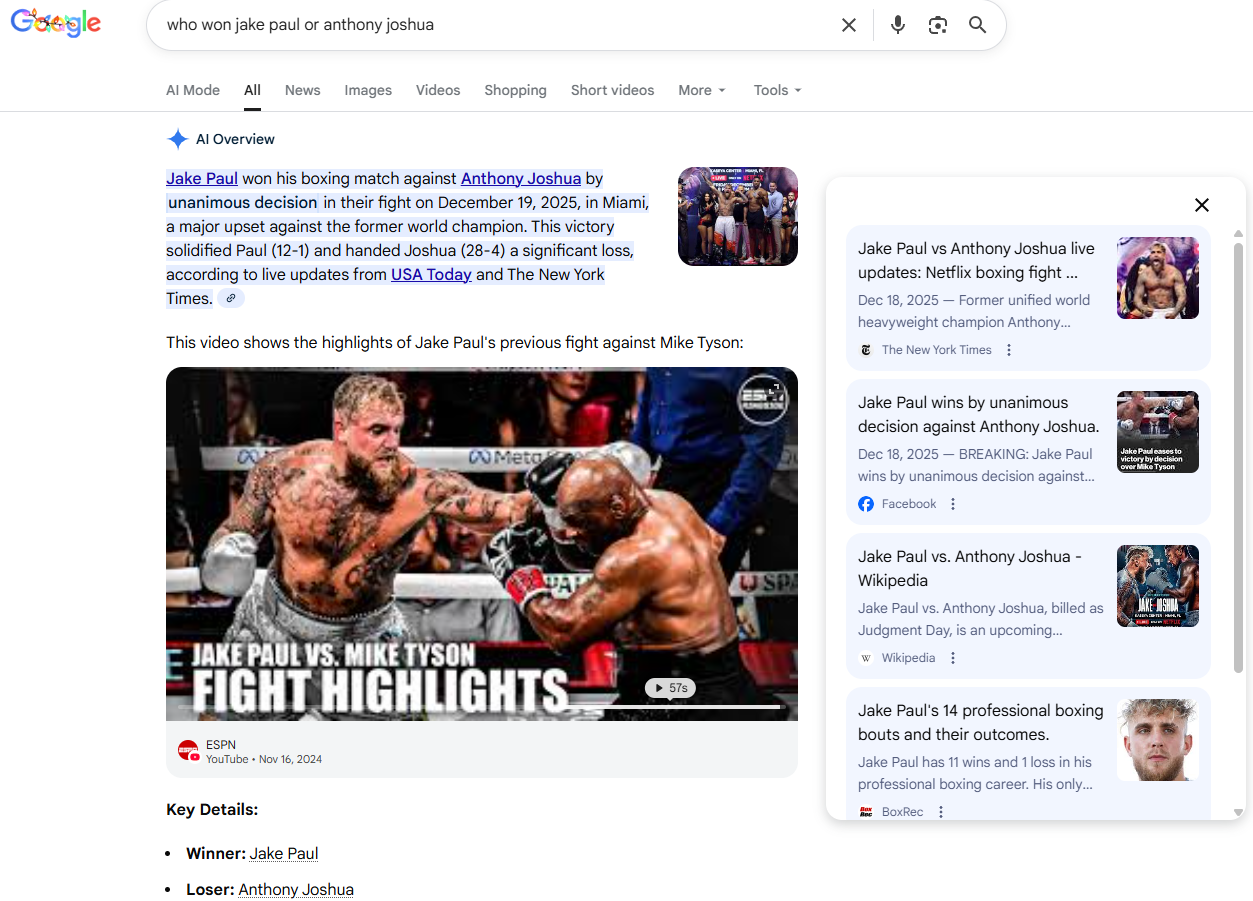}
    \caption{Misinformed AI Overview Response}
    \label{fig:misinformation}
\end{figure}

\textbf{Trending Queries}. We find that AIOs are rare for trending queries (8.1\%). The likelihood of an AIO being generated does increase as time passes:
AIOs are generated for 12.7\% of queries that began trending more than five days ago, compared to only 4.8\% of queries that started trending within the past five days. It is possible that the lack of AIOs represents a guardrail set in place because AIOs are susceptible to sourcing from misinformation when there is a lack of consensus across many sources. As a case study, we present a misinformed AIO response in Figure~\ref{fig:misinformation}. In anticipation of a boxing match later that night, the trending query ``who won jake paul or anthony joshua'' was incorrectly answered by the AIO, which declared Jake Paul as the winner. Interestingly, AIO cited several sources, despite only one of them (a Facebook post from a satirical sports account) claiming Jake Paul had won.

%

\section{Robustness Tests}
\label{sec:robustness}
To confirm that our findings are indicative of a meaningful disruption to the search engine industry, we perform a number of robustness tests.

\textbf{SerpAPI is representative of real user experiences}.\label{serpapi_robustness} The first test is to confirm whether our automated data collection methodology produces data that is representative of a real-user setting. Due to randomness in individual search engine results (see Section~\ref{subsec:consistency}), we cannot simply confirm that the exact same results are shown for queries submitted through SerpAPI and a real user's browser. Instead, we aim to show that the similarity between the API and manually retrieved sources is equivalent to the similarity between two API collections. The manual data collection occurs on one of the author's laptops from a signed-in Chrome profile, and the automated data is collected through SerpAPI on a desktop device from two cities. Data is simultaneously collected for the 100 queries in Section~\ref{subsec:consistency}. For the two API collections, the RBOs between two runs of AIO or SERP results are 0.67 and 0.81, respectively. The average RBO between manual and API collections is 0.67 and 0.79 for AIO and traditional search, respectively. These results indicate that SerpAPI yields source lists that are as close to a signed-in, manual browsing setup as repeated API collections are to each other, supporting SerpAPI as a representative proxy for real-user retrieval in our experiments.

One potential limitation to our approach is that real users that are signed into a Google profile may have higher consistency between two searches of the same query. For our two manual collections, we found average RBOs of 0.73 and 0.89 for AIOs and traditional search, respectively. These values are slightly higher than the similarities (0.69 and 0.86) computed in Table~\ref{tab:rbo_consistency} for the same device and location. Regardless, these results still show that AIOs are less consistent than traditional search, and show that our main result around internal search engine consistency holds for real users.       

\textbf{Robustness to Device and Location}. Given that device type and location influence the sources retrieved by SERP and AIO (see Table~\ref{tab:rbo_consistency}), we want to confirm that our findings from Sections~\ref{subsec:aio_generation},~\ref{subsec:rbo_jaccard_ndcg}, and~\ref{subsec:source_comparisons} hold under different settings. We collect two stratified random samples that are each 10\% of the full benchmark size. We collect data for these samples with the same method as our initial collection, except we change the SerpAPI device parameter to desktop and use different locations for each sample. Our major findings hold for desktop devices: AIOs appear on 67.17\% of all queries (52.2\% of ORCAS queries); AIOs are more frequent on longer, informational queries formatted as a question; and traditional search retrieves more sources from popular domains in comparison to AIOs and Gemini. The average RBO between AIO and SERP sources is higher on desktop than mobile (0.31 versus 0.23), but the main finding that source similarity is low between search engines holds true. 

\textbf{Comparison to Bing}. To contextualize the dissimilarity between Google's traditional and generative search engine results, we compare traditional Google Search with Bing, the second most popular search engine~\cite{stata_search_engine_market}. We use SerpAPI to collect organic search results from Bing and compare them to Google's organic search results for the 100 queries in Section~\ref{subsec:consistency}. We calculate an average RBO of 0.14 for the two search engines' retrieved sources, which is lower than the average RBO when comparing Google Search with either generative search engine. Although the retrieved results from Bing and Google Search are highly dissimilar, Bing's small market share has made this dissimilarity less important historically. For example, industry reports claim that the most popular Generative Search tool, ChatGPT, receives more daily search queries than Bing~\cite{vis_capitalist,first_page_sage,demand_sage_bing,demand_sage_gpt}, and even assuming that AIOs are generated on 50\% of users' Google searches would result in over 11 times the number of daily queries received by Bing. Thus, in comparison to Bing, publishers likely care more about their visibility in generative search results.

\section{Discussion: Societal Impact and Limitations}

\textbf{Considerations for Users}. 
From a user perspective, generative search engines offer convenience by providing integrated summaries alongside links for further reading, when needed. However, a major concern is that AI-generated information may be inaccurate. We find a clear example of this issue in our data (Figure~\ref{fig:misinformation}), and a recent news article from The Guardian highlights dangerously inaccurate AIO responses to health-related queries~\cite{guardian}. Due to the inherent challenges to deliver perfect solutions, users should use caution when leveraging generative search.

\textbf{Implication for Publishers}. Our results show that generative search will benefit niche content providers at the expense of more popular, established ones. Many large publishers have already taken actions against  generative AI companies by filing  copyright infringement lawsuits~\cite{times_suing,suing_65}, and claim that generative search had led to declining  traffic to their websites~\cite{ai_traffic_techcrunch,ai_traffic_forbes}. 

If generative search is here to stay, what strategies can publishers implement to adapt? Our results challenge the effectiveness of  GEO techniques~\cite{aggarwal_geo_2024, pfrommer-etal-2024-ranking,nestaas2025adversarial}. 
First, we find that the source lists produced by Gemini and the AIO are the least similar, compared with the pairs of source lists produced by traditional earch and either generative search engine (Table~\ref{tab:dataset_rbos}). 
This indicates the challenges of  performing well across multiple generative search engines. 
Second, we find that in comparison to traditional search, AIOs are less consistent between multiple runs of the same query (Table~\ref{tab:rbo_consistency}). 
As randomness has a large effect on whether a source  appears or is highly ranked in any given run of a query, optimization for high rankings in generative search may be unreliable.

Another decision that publishers need to make is whether to allow AI companies access to their content.
 Our findings in Figure~\ref{fig:regression} show that websites blocking the Google AI crawler are significantly less likely to be cited by generative search engines in comparison to traditional SERP. 
 While this is expected for Gemini, it is surprising that it also affects AIOs, given that Google only allows website to remove their content from AIOs by removing their content entirely from Google Search results~\cite{google_forces,google_forces_bloomberg,google_forces_sg,google_forces_medium}. Given the high presence of AIOs and the prominent position above the traditional search results, publishers may need to rethink their decision to block Google from accessing their content for AI training and grounding. 
%

\textbf{Implications for Retail and Service Websites.} We find that AIOs are generated infrequently for Amazon Retail queries (17.4\%) but frequently for Amazon Retail comparison and question queries (88.2\% and 92\%, respectively). This suggests that generative search plays a larger role during the consideration or research stages than the final purchase stage. AIOs are also moderately prevalent among the Localized queries subset (48.8\%), which is largely searches for service providers (e.g. ``cheapest restaurants near me''). 

Given that retrieved search results differ substantially between traditional and generative search engines for all aforementioned query sets, there may be an opportunity for lesser known providers of goods or services (i.e., those not ranking well in traditional search results) to optimize their content for visibility in generative search results. Although our results generally question the effectiveness of GEO techniques due to the dissimilarity in generative search engines' retrieved sources, prior research focusing on product and service queries has found that third-party reviews are heavily relied on by many generative search engines~\cite{chen2025products}. Thus, it may be beneficial for niche providers to invest in positive third-party reviews from sources frequently cited by generative search engines. 



%


\textbf{Implications to Generative Search}. 
We expect generative search engines to continue growing in popularity. However, important improvements are needed. First, inconsistency in responses from generative search engines for queries with the same intent undermines their usability and trustworthiness. Such inconsistency potentially pose serious risks for democracy  by presenting asymmetric information to different users~\cite{urman2021andwheresearchcomparative}. At a minimum, generative search engines should increase consistency across repeated runs of the same query and improve robustness to minor query variations, so that users seeking the same information are presented with similar content.

Second, future research should continue to address reliability issues in generative search, such as hallucinations and the spread of misinformation. 
Inaccuracies may stem from the reliance on less popular websites.  Although there is certainly an argument for using information from diverse sources, one concern is that the lesser known websites are less likely to be vetted for credibility and biases by third parties (e.g., MBFC).
Thus, to improve  quality, generative search engine companies should consider utilizing more information from popular and reputable sources, as traditional search engines do. Yandex appears to already implement this approach as their AI summaries are generated based on the  top five search results retrieved by its traditional search engine. 
%
%
In addition, generative search companies could consider funding independent evaluations of bias and credibility for niche websites, to promote source diversity without sacrificing reliability.

\textbf{Implications to the Ecosystem}.
%
%
Given the inherent imperfections of AI-based solutions, it is important to establish policies~\cite{euai2024} that regulate the use of generative search engines in contexts involving controversial issues (e.g., elected officials) or high stakes topics (e.g., identifying symptoms of a serious medical event).


Furthermore, we urge for the development of deals between digital publishers and AI companies~\cite{AI_publisher_deals} for a healthy online publishing and search ecosystem. Reduced traffic and the resulting decline in advertising revenues threatens publishers’ viability. In turn, a loss of high-quality content would ultimately undermine generative search engines themselves who may be ''starved'' of useful content for training and grounding. In addition to singular licensing deals, revenue frameworks could be established, such as negotiated licensing, pay-per-crawl, or other revenue-sharing arrangements, which compensate publishers for their content and enable generative search engines to utilize information from the most popular and reputable websites. Such a framework would better align the incentives of publishers and generative search providers, supporting a sustainable ecosystem where both sides benefit.

%
%

\yc{I like the discussions on regulation. I don't understand what "reciprocal access" refer to and delete it for now}
%
%

\textbf{Limitations}. Our experimental design and collected dataset lead to several limitations. First, this study solely focuses on Google, given its leading position in the search industry. To understand the generalizability of some results (e.g., generative search engine inconsistencies or preferences for niche content), it would be desirable to include other leading AI chatbots and generative search tools (e.g., ChatGPT, Perplexity AI, and Bing Copilot). Inclusion of open-source generative search tools would allow for better testing of the underlying mechanisms that contribute to low internal consistency and preference for niche sources. Second, although it was necessary to collect a large volume of data, reliance on APIs for data collection does limit our results in that we cannot study how generative search outputs vary by user characteristics (beyond device and city). It would be interesting for future user studies to explore how the users' characteristics (browsing as a signed-in user) impact the prevalence of AIOs and the generative search engines' retrieved sources and text summaries.

Finally, our analyses are both descriptive in nature and limited to a single point in time. Given these limitations, this study should motivate future research that considers the broader ecosystem implications, beyond just the user impact, of generative search systems. Generative search is changing quickly, and it is our hope that providing a public benchmark dataset of queries and returned URLs will enable future research to document changes in generative search engine behavior. 

\section{Conclusions}
We conducted a large-scale comparison of the results generated by traditional Google Search (SERP), Google AI Overviews (AIO), and Gemini for 14{,}212 total queries (including 11,500 time-invariant queries released in our benchmark dataset). There are several key findings. AIOs are generated for a substantial share of real-user searches and appear above organic results. Across search engines, the cited sources differ substantially, leading to meaningfully different source exposure for users and websites. Generative search shifts visibility away from popular and institutional domains and toward Google-owned properties. Surprisingly, AIOs retrieve fewer sources from websites blocking the Google-Extended bot, despite having access, which may lead some publishers to rethink their blocking strategies. Finally, we find that generative search engines are less stable than SERP across runs and more sensitive to changes in device type and minor query edits. Generative search's impact on users is most concerning for high-stakes queries, where AIOs frequently appear and may rely on low-credibility sources or adopt a stance. We release our benchmark query set, processed data, and code to support replication and longitudinal monitoring of generative search behavior. Future work may consider how personalization plays a role in the results retrieved by generative search systems or conduct user studies to better understand how generative search citations impact publishers' traffic and advertising revenues.

\yc{if have space, we can add 1-2 sentences about implication and/or 1 sentence about future work}

\begin{acks}
Research reported in this publication was supported in part by the NSF under Grant No. CNS2237328, the National Center For Advancing Translational Sciences of the National Institutes of Health under Award Number UM1TR004789,  as well as by the Martin Tuchman’62 Chair Endowment and the Leir Foundation. The content is solely the responsibility of the authors and does not necessarily represent the official views of the funders.
\end{acks}

\bibliographystyle{ACM-Reference-Format}
\balance
\bibliography{sigir_submission}




\end{document}